\documentclass[runningheads]{llncs}
\usepackage{graphicx}
\usepackage{bbding}
\begin{document}
\title{When Diffusion MRI Meets Diffusion Model: A Novel Deep Generative Model for Diffusion MRI Generation}
\titlerunning{When Diffusion MRI Meets Diffusion Model}
%
\author{Xi Zhu\textsuperscript{1}\and
Wei Zhang\textsuperscript{1} \and
Yijie Li\textsuperscript{1} \and 
Lauren J. O'Donnell\textsuperscript{2} \and 
Fan Zhang\textsuperscript{1}\textsuperscript{(\Envelope)}}
\authorrunning{Xi Zhu et al.}
\institute{University of Electronic Science and Technology of China, Chengdu, China \and
 Harvard Medical School and Brigham and Women's Hospital, Boston, USA\\
\email{fan.zhang@uestc.edu.cn}\\}
\maketitle              
\begin{abstract}
Diffusion MRI (dMRI) is an advanced imaging technique characterizing tissue microstructure and white matter structural connectivity of the human brain. The demand for high-quality dMRI data is growing, driven by the need for better resolution and improved tissue contrast. However, acquiring high-quality dMRI data is expensive and time-consuming. In this context, deep generative modeling emerges as a promising solution to enhance image quality while minimizing acquisition costs and scanning time. In this study, we propose a novel generative approach to perform dMRI generation using deep diffusion models. It can generate high dimension (4D) and high resolution data preserving the gradients information and brain structure. We demonstrated our method through an image mapping task aimed at enhancing the quality of dMRI images from 3T to 7T. Our approach demonstrates highly enhanced performance in generating dMRI images when compared to the current state-of-the-art (SOTA) methods. This achievement underscores a substantial progression in enhancing dMRI quality, highlighting the potential of our novel generative approach to revolutionize dMRI imaging standards.

\end{abstract}
\section{Introduction}

Diffusion MRI (dMRI) is an advanced neuroimaging tool to characterize the underlying brain tissue microstructure~\cite{basser1994mr} and is widely used for studying the brains~\cite{pannek2014magnetic,zhang2022quantitative}. Currently, there is an increasing interest in high-quality dMRI data for better resolutions and enhanced tissue contrast such as dMRI data from a 7T scanner~\cite{chilla2015diffusion,sotiropoulos2016fusion,ramos2020high,vu2015high}. However, acquiring such high-quality dMRI data necessitates advanced MRI scanners and/or acquisition protocols, which are not always accessible and thus remain impractical in real-world applications.

Generation of dMRI using machine learning offers high promise to improve image quality while reducing acquisition costs and scanning time. This task generally involves image-to-image translation to learn a mapping from low-quality (source) to high-quality (target) data, which can subsequently predict (or generate) high-quality data when only low-quality data is available. Traditional methods have used techniques such as random forest~\cite{alexander2014image,nedjati2014machine} to map voxel patches from low-quality to high-quality data. With the advances in deep learning, many studies have used deep networks for dMRI generation~\cite{tanno2017bayesian,cetin2018harmonizing,hirte2021realistic,jha2023trganet}. For instance, Karayumak et al. introduced a convolutional neural network (CNN) approach~\cite{cetin2018harmonizing} and Ranjan Jha et al. employed a more sophisticated generative-adversarial network (GAN) approach~\cite{jha2023trganet} to generate high-quality dMRI data from 3T to 7T.

Recently, diffusion models have demonstrated remarkable results for generative modeling in medical imaging~\cite{kazerouni2023diffusion}, which may provide a powerful tool for dMRI data generation. In brief, a diffusion model comprises a forward diffusion stage, where input data is progressively perturbed by Gaussian noise, followed by a reverse diffusion stage aimed at gradually reverting the process to recover the original input. The Denoising Diffusion Probabilistic Model (DDPM) \cite{ho2020denoising} is one representative diffusion model for image generation. Many variations of DDPM have been proposed. For example, the same research group of DDPM introduced the concept of classifier-free guidance~\cite{ho2022classifier} to remove complex classifiers and make the model simpler. The Denoising Diffusion Implicit Model (DDIM) \cite{song2020denoising} skips some steps to accelerate the sampling speed in DDPM. The Latent Diffusion Model (LDM)~\cite{rombach2022high} made DDPM work on latent space, which can handle high-resolution images to increase computational efficiency. Currently, diffusion models show great advantages in medical imaging, such as anomaly detection~\cite{wolleb2022diffusion}, signal reconstruction~\cite{ozturkler2023smrd}, and image generation~\cite{pinaya2022brain}. In the dMRI field, one recent study has successfully used the diffusion model for data denoising~\cite{xiang2023ddm}. Yet, there is no work for dMRI generation using diffusion models.

The application of diffusion models for dMRI generation is a challenging task due to the uniqueness of dMRI data. First, dMRI is a unique, multi-dimensional image dataset that describes not only the strength but also the orientation of water diffusion. Applying diffusion models to high-dimensional data presents significant challenges, leading recent research to focus primarily on slices or single volumes. This approach overlooks the crucial 3D orientation information offered by dMRI, which is essential for understanding the complex spatial relationships and orientations in the data. Secondly, training diffusion models for data quality enhancement necessitates the availability of both standard and high-quality data, such as datasets from 3T and 7T scanners. However, acquiring high-quality data is challenging, often resulting in limited data to restrict the potential for using generative models for a large scale data analysis.

In this paper, we present a novel generative approach for dMRI generation using deep diffusion models. To the best of our knowledge, this is the first work to apply diffusion models specifically for enhancing the quality of dMRI data. Our method has the following contributions: 1) proposing using LDM for the high-quality 7T rotation invariant spherical harmonic (RISH) features generation and reconstructing the 4D dMRI data, 2) designing a transfer learning strategy for autoencoder training to address the scarcity of high-quality 7T data, and 3) a super-resolution module to remove resolution differences. We demonstrated our method through an image mapping task aimed at enhancing the quality of dMRI images from 3T to 7T. Our approach demonstrates highly enhanced performance in generating dMRI images when compared to several compared methods, indicating a notable advancement in dMRI quality improvement.

\section{Method}

\begin{figure}[!t]
\includegraphics[width=\textwidth]{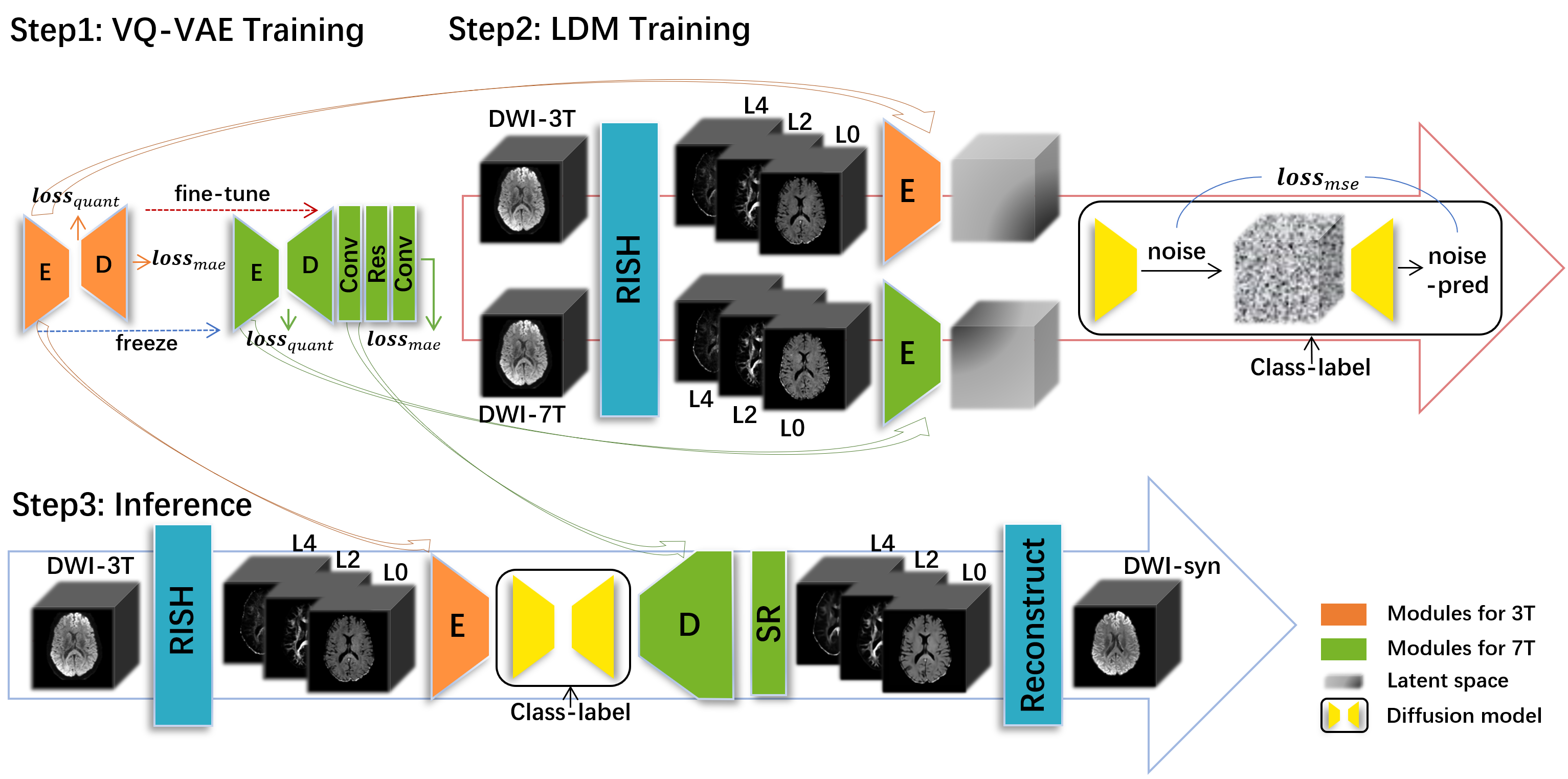}
\caption{Overview of the proposed method.} \label{fig1}
\vspace{-0.3cm}
\end{figure}

Fig.~\ref{fig1} gives an overview of our method. First, RISH features~\cite{mirzaalian2016inter,karayumak2019retrospective} are computed for an efficient and compact representation of the input 3T and 7T dMRI data. Next, we train a deep generative model to learn the RISH features of 7T. This step includes: 1) two autoencoders that, respectively, learn latent features of the 3T and 7T RISH features, where we design a fine-tuning strategy to address the scarcity of high-quality 7T training data; and 2) a classifier-free guidance DDPM to generate 7T-like latent features from 3T, where we introduce a super-resolution module to enable simultaneous dMRI signal generation and spatial resolution enhancement. Finally, during inference, the RISH features of a testing 3T dataset are encoded into the latent space using the 3T encoder, followed by a DDIM process to generate 7T-like RISH features for reconstructing a high-quality 7T dMRI dataset. 

\subsection{dMRI Datasets}

We used the dMRI data provided in the Human Connectome Project (HCP)~\cite{van2013wu}. In total, data from 1065 subjects were used, of which 171 had both 3T and 7T dMRI data, and 894 had only 3T data. The acquisition parameters of 3T dMRI data were: $TE=89.5ms$, $TR=520ms$, and voxel size=1.25$\times$1.25$\times$1.25mm$^3$, 18 baseline images, and 270 diffusion-weighted images distributed evenly at $b=1000/2000/3000\ s/mm^3$; and those of the 7T data were: $TE=71.2ms$, $TR=7000ms$, and voxel size=1.05$\times$1.05$\times$1.05mm$^3$, 15 baseline images, and 128 diffusion-weighted images distributed evenly at $b=1000/2000\ s/mm^3$. The provided dMRI data has been preprocessed as in~\cite{glasser2013minimal}. In our study, for simplicity, we used only the single shell $b=1000$ data in both 3T and 7T data. 

\subsection{dMRI signal representation and reconstruction using RISH}

In dMRI, the signal S of each voxel can be represented in a basis of spherical harmonics (SH)~\cite{descoteaux2007regularized}: $S\approx\sum_i\sum_jC_{ij}Y_{ij}$, where $C_{ij}$ is the coefficient of SH basis function $Y_{ij}$ at order $i$ and degree $j$. Then, from the SH coefficients, the RISH features at each order $i$ can be computed as follows:
\begin{equation}
||C_i||^2=\sum_{j=1}^{2i+1}(C_{ij})^2
\end{equation}
One of the benefits of the RISH features is that they can be appropriately scaled to modify the dMRI signals without changing the principal directions of the fibers~\cite{mirzaalian2016inter}. In addition, the RISH features give a compact and uniform representation of the dMRI data regardless of the number of gradient directions. In our study, we computed the RISH features for each subject’s 3T and 7T images with SH orders of $i = \{0, 2, 4\}$ as suggested in~\cite{de2022cross}.

For dMRI data generation from 3T to 7T, during training stage (see Section 2.3), we can learn 7T-like RISH features by computing so-called scale maps between the two datasets, as:
\begin{equation}
\lambda_i=\sqrt{\frac{\|C_i\|_{7T}^2}{\|C_i\|_{3T}^2+\tau}}
\end{equation}
where $\tau$ is a constant with a very small value. Then, during inference when only 3T data is available, the scale maps can be predicted via the learned models, which can be subsequently applied to the SH coefficients from 3T images to generate 7T-like RISH features, as follows:
\begin{equation}
\hat{C}_{ij}=\lambda_iC_{ij}
\end{equation}
where $\hat{C}_{ij}$ is the predicted SH coefficients of 7T. Finally,  a high-quality 7T dMRI dataset can be generated by reconstructed dMRI signals through Eq. (1) with the predicted  $\hat{C}_{ij}$ and  the SH basis function $Y_{ij}$. 

\subsection{Latent diffusion model}

To fully leverage the 3D properties of the dMRI RISH features and tackle the issue of limited availability of high-quality dMRI data, we propose a new architecture based on Latent Diffusion Models (LDM) complemented by a fine-tuning strategy. In detail, we use the Vector Quantised-Variational AutoEncoder (VQ-VAE)~\cite{van2017neural} to compress the whole brain image to the latent space. VQ-VAE quantifies the latent representation of images to get a better latent presentation.  To address the differences between 3T and 7T MRI datasets effectively, we train two separate VQ-VAE models, one for each dataset type.  However, the VQ-VAE model struggles to produce satisfactory outcomes for 7T data due to the limited volume of available high-quality data. To overcome this limitation, we introduce the application of transfer learning. We first train a model extensively on the abundant 3T dataset and subsequently fine-tune this model using 7T data. Both of the models use MAE loss and quantization loss during the training process.

Then, the data generation via the diffusion process is performed in the latent space and takes the output $x$ of the VQ-VAE’s encoder as input. The process can be generally divided into two parts: the forward noising process and the backward denoising process. The forward noising process $q$ is defined as follows:
 \begin{equation}
     q(x_t|x_{t-1})=N(x_t;\sqrt{1-\beta_t}x_{t-1};\beta_tI)
 \end{equation}
 Where $x_t$ is the noisy latent features that are obtained by an iterative process of noise addition, $\{\beta_{1},\beta_{2},...,\beta_{t},...,\beta_{T}\}$ is a series constant, and $t\in\{0,...,T\}$ is a moment during the noise addition process. Noisy features at the moment t can be written as:
\begin{equation}
x_{t}=\sqrt{\overline{\alpha}_{t}}x_{0}+\sqrt{1-\overline{\alpha}_{t}\epsilon},\ \mathrm{with}\ \epsilon\in N(0,I)
\end{equation}
with $\alpha_{t}=1-\beta_{t}$ and $\overline{\alpha}_{t}=\prod_{s=1}^{t}\alpha_{s}$. The denoising process $p_\theta$ relies on a U-Net to predict $x_{t-1}$ from $x_t$ by optimizing the U-Net’s parameters $\theta$. It can be given as
\begin{equation}
    p_\theta(x_{t-1}|x_t)=N(x_{t-1};\mu_\theta(x_t,t);\sum_\theta(x_t,t))
\end{equation}
 The U-Net can be denoted as $\epsilon_{\theta}$, and MSE loss is used to train this model, as:
\begin{equation}
    L=||\epsilon-\epsilon_{\theta}(\sqrt{\overline{\alpha}_{t}}x_{0}+\sqrt{1-\overline{\alpha}_{t}\epsilon})||_{2}^{2},\ \mathrm{with}\ \epsilon\in N(0,I)
\end{equation}
  In the sampling process, we can encode the latent features by adding noise based on the model’s output at step $t$:
  \begin{equation}   x_{t+1}=x_{t}+\sqrt{\overline{\alpha}_{t+1}}[(\sqrt{\frac{1}{\overline{\alpha}_{t}}}-\sqrt{\frac{1}{\overline{\alpha}_{t+1}}})x_{t}+(\sqrt{\frac{1}{\overline{\alpha}_{t+1}}-1}-\sqrt{\frac{1}{\overline{\alpha}_{t}}-1})\epsilon_{\theta}(x_{t},t)]
  \end{equation}
In the above diffusion model, the class labels (3T and 7T) are used to control the direction of diffusion model's generation, they can be encoded to class-embeddings and introduced to the U-Net backbone through the cross-attention mechanism implementing $Attention(Q,K,V)=softmax\left(\frac{Q K^{T}}{\sqrt{d}}\right) \cdot V$, with 
\begin{equation}
    Q=W_{Q}^{(i)} \cdot \varphi_{i}\left(z_{t}\right), K=W_{K}^{(i)} \cdot \tau_{\theta}(c), V=W_{V}^{(i)} \cdot \tau_{\theta}(c).
\end{equation}
with $c$ is the class label, $\tau_{\theta}$ represents a specific encoder that encodes $c$ to embedding and $\varphi_{i}$ denotes a intermediate representation of the U-Net.

The generation of controlled diffusion models is divided into the sum of an unconditional generation process $\epsilon_\theta(x_t)$ and a conditional generation process $\epsilon_\theta(x_t,c)$. So, we turn certain labels into uncertain ones with a probability which is set to a hyperparameter in the training process. In the inference process, we turn the 3T latent features into the 7T features with class-embedding guidance and the U-Net’s prediction can be given as:
\begin{equation}
    \bar{\epsilon}_\theta(x_t,c)=(1+\omega)\epsilon_\theta(x_t,c)-\omega\epsilon_\theta(x_t)
\end{equation}
the $\omega$ represents the guidance scale of class embedding.

Finally, we use the dataset generated by LDM to train the super-resolution module located at the end of 7T VQ-VAE. The architecture of it includes two convolution layers and  middle residual layers. We apply a similar training strategy with SR-CNN and use the MSE loss to optimize module.

\section{Experimental Comparisons}

We compare the performance of the proposed method with CNN-based ~\cite{cetin2018harmonizing} and GAN-based network architecture ~\cite{rombach2022high}. The CNN-based method designed a deep 3D convolutional network, and the GAN-based method contained an autoencoder with an attention mechanism trained by an adversarial framework. These methods also used 4 orders of RISH features as input and reconstruction of each method was the same. The synthesis quality was evaluated using normalized mean squared error (NMSE), and structural similarity index (SSIM) across multiple scales. 17 subjects with both 3T and 7T data were randomly selected and left for testing, while the remaining were used for training and validation of the autoencoders and the DDPM. There were 1065 subjects’ data for 3T VQ-VAE training and 171 7T data for fine-tuning 7T VQ-VAE. Finally, 342 subjects both having 7T and 3T data were used to train LDM. All of these datasets were divided into training sets and test sets at a ratio of $9:1$, and every test set included 17 test subjects. All metrics are calculated over 3D volumes to ensure comprehensive analysis.

The implementations of VQ-VAE and LDM were done using Pytorch ~\cite{imambi2021pytorch} and MONAI ~\cite{pinaya2023generative} framework. All the RISH features were downsampled to $96\times96\times96$ before inputting to the LDM, and so were other test methods. For the hyperparameters,  we set the number of embedding dimensions to 32 and the number of embeddings to 256 for training VQ-VAE. As for LDM, we choose  different guidance scales and levels of noise addition for RISH features of each order during sampling steps. We used the AdamW optimizer with a learning rate of $1\times10^{-4}$ and set training epochs to 200 and 1000 for VQ-VAE and LDM respectively. For the detailed architecture, we used attention heads at the third and fourth layers of U-Net used in LDM to predict noise. All computation was conducted on RTX 3090 GPUs.

\begin{table}[!t]
\centering
\caption{Comparison of NMSE and SSIM in RISH and FA across different methods.}
\label{tab1}
\begin{tabular}{lllll}
\hline
NMSE$\downarrow$:&RISH\_L0&RISH\_L2& RISH\_L4 & FA \\
\hline
CNN & $0.126\pm0.014$&$0.143\pm0.011$ & $0.495\pm0.107$ & $0.053\pm0.007$ \\
GAN  &$0.129\pm0.029$ & $0.427\pm0.051$ & $1.652\pm0.360$ & $0.118\pm0.009$ \\
Diffusion &$\mathbf{0.105\pm0.026}$ & $\mathbf{0.102\pm0.017}$ & $\mathbf{0.158\pm0.031}$ & $\mathbf{0.044\pm0.008}$ \\
\hline
 SSIM$\uparrow $: & & & & \\
\hline
 CNN &  $0.889\pm0.008$ & $0.959\pm0.006$ & $0.956\pm0.016$ & $0.958\pm0.006$ \\
 GAN &  $0.915\pm0.012$ & $0.893\pm0.010$ & $0.943\pm0.004$ & $0.902\pm0.010$ \\
 Diffusion & $\mathbf{0.922\pm0.009}$ & $\mathbf{0.961\pm0.007}$ & $\mathbf{0.967\pm0.002}$ & $\mathbf{0.966\pm0.007}$ \\
\hline
\end{tabular}
\end{table}

 \begin{figure}[!t]
\includegraphics[width=\textwidth]{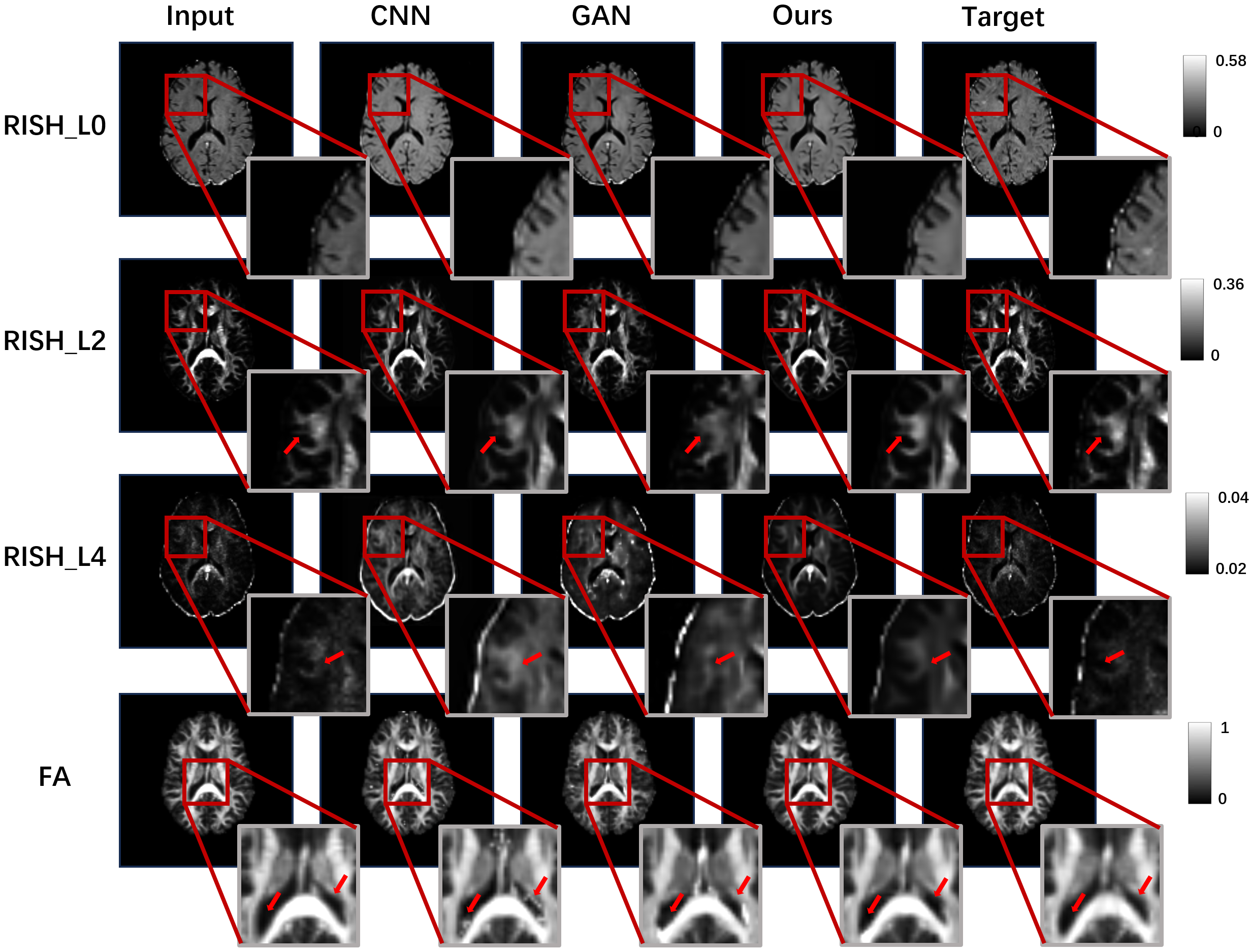}
\caption{Results for the RISH features and FA generated by different methods. } \label{fig2}
\end{figure}

\section{Results}

 \subsubsection{SOTA comparison.} 
\begin{figure}[!t]
\includegraphics[width=\textwidth]{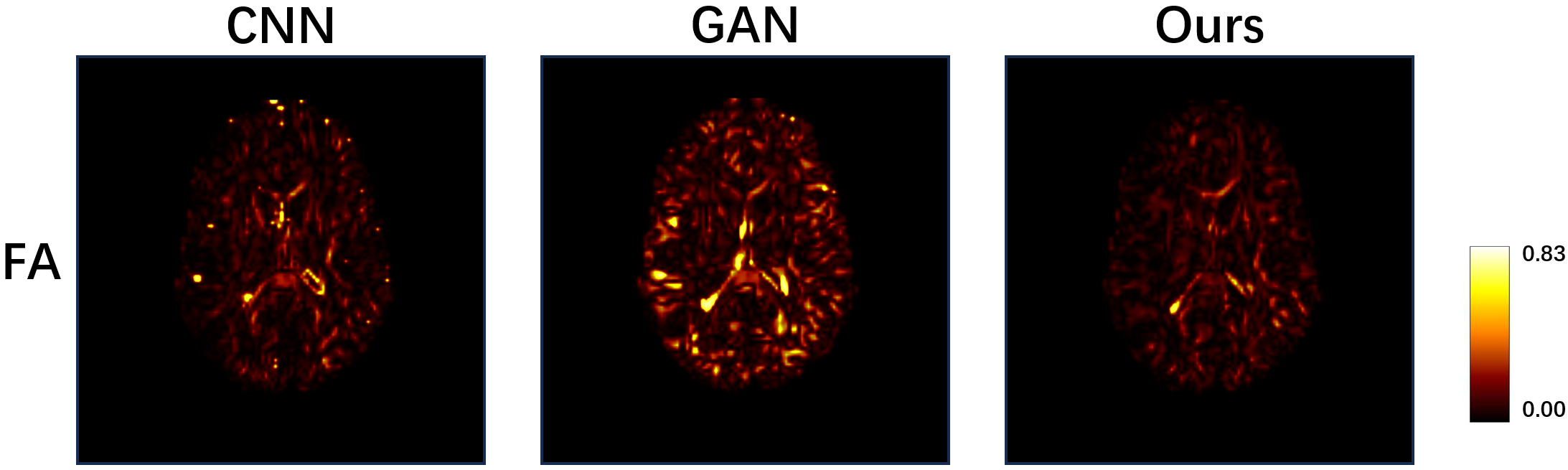}
\caption{Difference maps of FA images.} \label{fig3}
\end{figure}

Table ~\ref{tab1} gives the mean NMSE and SSIM for the RISH features and the FA images, where we can see that our method outperforms the other two methods in quantitative metrics. Fig. ~\ref{fig2} provides a visual comparison of various RISH features produced by the different methods and the FA images against the target data. Our approach is distinguished by producing images that more closely resemble the ground truth than the other compared methods. The CNN-based method tends to uniformly increase the intensity of all voxels across the input images, leading to a loss of contrast information between different regions. Meanwhile, the GAN-based method fails to preserve some of the structural details in higher-order L2 and L4 RISH features. Figure ~\ref{fig3} illustrates the difference maps of FA between the predicted and target data, demonstrating that our method achieves the closest resemblance to the target data, further highlighting its accuracy in generating high-quality dMRI images.

\subsubsection{Ablation experiment.} 
To explore the effects of the fine-tuning and the super-resolution module proposed in our method, we conducted ablation studies from these two aspects. We adopted a network the same as the 3T VQ-VAE and trained it on the collection of the 7T datasets. The DWI data were reconstructed with the same process which excluded the super-resolution module, and then compared the FA images with the NMSE. For the experiments examining the super-resolution module, we first applied B-spline interpolation to upsample the 3T data to the same resolution as 7T and register it to the 7T space. Following this, we enhance 3T data with our method and data acquired by 7T scanner. Table. 2 presents the ablation study results. We can observe a notable improvement in performance post-fine-tuning. Moreover, the introduction led to a further reduction in the NMSE of the FA images.

\begin{table}[!t]
\centering
\caption{Ablation study results}\label{tab2}
\begin{tabular}{llll}
\hline
 Fine-tuning\ \ & Super-resolution\ \  &  NMSE$\downarrow$\ \ & SSIM$\uparrow$ \\
\hline
   -  & -  &  $0.046\pm0.008$ & $0.962\pm0.008$\\
   \Checkmark  & -  &  $0.044\pm0.008$ & $0.966\pm0.007$ \\
   \Checkmark  & \Checkmark  &  $\mathbf{0.042\pm0.004}$ & $\mathbf{0.967\pm0.007}$\\
\hline
\end{tabular}
\end{table}

To investigate the impact of the fine-tuning and super-resolution components within our method, we carried out  the following ablation studies. First, we utilized a network identical to the 3T VQ-VAE and trained it exclusively with the 7T dataset without fine-tuning. Second, we performed a method without using the super-resolution module, instead using a B-spline interpolation to upscale the 3T data to match the resolution of 7T data. The same quantitative measures NMSE and SSIM  were used for experimental comparison. Table ~\ref{tab2} shows the comparison results, showing a significant improvement in using the fine-tuning process and the super-resolution module.

\section{Conclusion}

We present a novel framework that leverages the latent diffusion model and rotation invariant spherical harmonic to generate high-quality dMRI data. We applied the proposed method for image generation on the HCP dataset and successfully generated the 7T-like dMRI image from 3T. Our method largely outperforms current SOTA methods in generating dMRI images, marking a major advancement in dMRI quality enhancement. This underscores the potential of our innovative generative method to transform dMRI imaging standards.

\begin{credits}
\subsubsection{\ackname} This work is in part supported by the National Key R\&D Program of China (No. 2023YFE0118600), the National Natural Science Foundation of China (No. 62371107), and the National Institutes of Health (R01MH125860, R01MH119222, R01MH132610, R01NS125781).  

\subsubsection{\discintname}
The authors have no competing interests to declare that are
relevant to the content of this article.
\end{credits}
%
%
%
%
\bibliographystyle{splncs04}
\bibliography{paper-2842.bib}
\end{document}